\pdfoutput=1 
\documentclass{JINST}

\title{Precision Astronomy with Imperfect Fully Depleted CCDs -- 
An Introduction and a Suggested Lexicon}

\author{Christopher W. Stubbs$^a$,\thanks{Corresponding author.}~\\
\llap{$^a$}Department of Physics \& Department of Astronomy, Harvard University,\\
17 Oxford Street, Cambridge MA, USA 02138\\
E-mail: \email{stubbs@physics.harvard.edu}}

\abstract{This paper summarizes the introductory presentation for a workshop (held Nov 18,19 2013 at Brookhaven National 
Laboratory) that explored the challenges associated with making precision astronomical measurements using deeply depleted 
= ``thick" =``high-$\rho$'' CCDs.
While thick CCDs do provide definite advantages in terms of increased quantum efficiency at wavelengths 700~nm~$<\lambda<$~1.1 $\mu$m
and reduced fringing from atmospheric emission lines, these devices also exhibit undesirable features that pose a challenge to precision determination 
of the positions, fluxes, and shapes of astronomical objects, and for the precision extraction of features in astronomical spectra. 
For example, the assumptions of a perfectly rectilinear pixel grid and of an intensity-independent point spread 
function become increasingly invalid as we push to higher precision measurements. 
Many of the effects seen in these devices arise from lateral electrical fields within the detector, that produce charge transport anomalies that have been
previously misinterpreted as quantum efficiency variations. Performing simplistic flat-fielding therefore {\em introduces} systematic errors
in the image processing pipeline. One measurement challenge we face is devising a combination of calibration methods and algorithms that can distinguish genuine quantum 
efficiency variations from charge transport effects. These device imperfections also confront spectroscopic
applications, such as line centroid determination for precision radial velocity studies. Given the scientific benefits of improving 
both the precision and accuracy of astronomical measurements, we need to identify, characterize, and overcome these various detector 
artifacts. In retrospect, many of the detector features first identified in thick CCDs also afflict measurements made with more 
traditional CCD detectors, albeit often at a reduced level since the photocharge is subject to the perturbing influence of lateral electric fields for a shorter time interval. 
I provide a qualitative overview of the physical effects we think are responsible for the observed 
device properties, and provide some perspective for the work that lies ahead. Finally, I take this opportunity to make a plea for establishing a 
clear and consistent vocabulary when describing these various detector features, and make some suggestions for a standard lexicon based on 
discussions at the workshop. A more refined understanding of the device imperfections we are working to circumvent lies ahead, and this workshop was convened to help us find our way. }


\begin{document}

\section{Introduction: improving the precision of astronomical measurements requires understanding, characterizing, and overcoming detector artifacts}\label{sec:intro} 

Detectors lie at the heart of the astronomical instruments that have led us into a golden age of astronomy, astrophysics, and cosmology. 
The November 2013 workshop on ``Precision Astronomy with Fully Depleted CCDs'' brought together experts in CCDs, algorithms, and 
science to discuss how to best use these devices to address forefront topics, and in particular how to overcome 
CCD features that limit the measurement precision we can presently achieve. 

At ``optical'' wavelengths, 300 nm $<\lambda<$ 1.1 microns, CCDs remain the sensor of choice for most astronomical instrumentation applications. As interest 
in the near-IR regime has increased, many projects have selected thick, fully depleted silicon CCDs. These devices, 
with current generation parts having thicknesses ranging up to 250 $\mu$m, have increased quantum efficiency at near-IR wavelengths. 
But they also come with some downsides, which impact our ability to extract Poisson-limited results from our observations. 

The parameters of interest for astronomical imaging observations include the positions (astrometry), fluxes and colors (photometry), and shapes of 
celestial sources, and for spectroscopic observations we need to measure line centroids, equivalent widths, and spectrophotometry.  
Detector artifacts that arise in thick CCDs do present challenges to our ability to measure these properties of interest. 

Improving both the accuracy and precision of astronomical measurements is highly motivated. 
A sampling of illustrative topics on the ``precision frontier'' of astrophysics that face challenges from imperfect detectors include:

\begin{itemize}
\item{} The use of type Ia supernovae for probing the nature of dark energy is presently impeded by systematic errors in precision photometric calibration, 
and we need to make relative flux measurements that are reliable at the sub-percent level. This in turn requires a flat fielding procedure that is 
not contaminated by systematic errors arising from charge transport effects within the detectors.
\item{} The determination of shapes of galaxies for weak lensing studies requires a detailed understanding of the point spread function (PSF) of the instrument, 
and subtle detector features can impact the spatial variation of the PSF across the field. Moreover, there is now compelling evidence that the 
shape of the PSF depends on the intensity of the source, and this undermines the common assumption that the PSF for a bright 
star is just a scaled version of the PSF of a fainter star, with the same FWHM. As shown at this workshop, this intensity-dependent PSF is an intrinsic property of 
essentially all CCDs in Nyquist-sampled imaging systems, and is in addition to the inevitable contribution from non-linearities in the analog signal chain.  
This intensity-dependent PSF distortion within the CCD is a consideration for both ground-based and space-based next-generation weak lensing projects. 
Also, the distortions in the mapping function within the detector from the sky onto the pixel grid can be a source of error for shape measurements across a stack of images. 
\item{} Radial velocity searches for extra-solar planetary systems require the detection of subtle shifts in high resolution spectra. 
Again, both the flat fielding considerations and the potential for misinterpretations due to pixel grid distortions can limit measurement precision. 
\item{} Photometric searches for transit occultations require sub-part-per-thousand photometric precision. Some projects have achieved remarkable (200 ppm!)
precision \cite{WASP}  by defocusing and 
spreading the flux of an isolated bright source over a wide region on the CCD array, 
essentially averaging out the short-spatial-scale flatfielding errors discussed below. But this approach won't work for faint or crowded sources.  
\end{itemize}

The simple fact that most astronomical measurements do not reach the measurement floors established by quantum mechanics
and by thermodynamics (namely Poisson photon statistics in former case and read noise and dark current in the latter) indicates that we
have considerable remaining scientific opportunity ahead of us as we push towards higher precision.   Sensor imperfections are among the impediments to our achieving
these fundamental measurement limits. Some of these unwanted CCD features are enhanced in thick CCDs (mainly effects 
arising from lateral electric fields) while other undesirable features (particularly fringing) are suppressed relative to thinner, more conventional devices.
Some of the detector ``features" of concern to us are:

\begin{itemize}
\item{} Image persistence, where deferred/trapped charge from bright sources is manifest in subsequent images,
\item{} Trapped holes that can produce variable streaking features in images, 
\item{} Loss of full well depth when exposed to high light levels while under bias,
\item{} Intensity-dependent point spread functions, 
\item{} Apparent variation in flat field response at the edges of sensors (``edge distortion"),
\item{} Anomalies in photon transfer curves, that appear to violate the statistical behavior one would expect from Poisson noise in uncorrelated pixels, 
\item{} Astrometric perturbations that introduce distortions in the mapping between sky and pixel coordinates on various spatial scales,  
\item{} Differences in subtle device characteristics that are different in the row and column directions, and
\item{} Both photometric and astrometric residuals that correlate with radially symmetrical ``tree rings'' seen in flat-fields, arising from impurity gradients in the silicon. 
\end{itemize}

Many if not all of these effects are seen (sometimes only in retrospect, see \cite{Astier}) in more conventional thin CCDs. In fact one might argue that none of them 
are actually new. Quoting from a 1995 paper that describes the calibration of the Wide Field/Planetary Camera on HST \cite{WFPC}: 
{\em ``There appears to be a systematic error in photometry as a function of the centering along
columns at the $<$2\% level; for observations of many objects, this effect would contribute $\le$1\% rms scatter. Any similar effect along 
rows is at a significantly smaller amplitude".} I make no attempt here to provide a comprehensive review of previous work, but illustrative examples of 
papers that discuss departures from the rectilinear array idealization include references \cite{Kotov} and \cite{Smith}.  Some initial results on the 
intensity-dependent PSF effect appears in \cite{Astier}. In particular I want to highlight Mark Downing of ESO who has for some time
been drawing many of these anomalies to the attention of the astronomical instrumentation community \cite{Downing}. 

But the growing impetus to fully understand and compensate for these sensor imperfections {\em is} new, and is largely motivated by the desire to achieve better
precision and accuracy in astronomical measurements. One aspect we now recognize as important is 
properly interpreting the pixel-to-pixel variation seen in flatfields images. Some of this variation results from genuine variation in quantum efficiency (the 
likelihood of a photon converting into a collected photoelectron (or photo-hole)), whereas some aspects of the pixel-to-pixel variation comes about purely from 
charge transport effects within the silicon.  It is essential that we distinguish between quantum efficiency variations and charge transport effects, which can 
(as discussed below) come about from diverse physical phenomena. 

I will close this introductory section with some intentionally provocative assertions. 

\begin{enumerate}

\item{} The astronomical instrumentation community has misinterpreted flatfields for decades, and have in many (most?) cases been introducing errors that limit measurement precision. Na\"{\i}ve flatfielding is a bad thing. In addition to errors introduced by our inability to distinguish stray and focused light paths in 
flats taken with diffuse surface brightness sources (from the dome, twilight or night sky), using surface brightness to perform flux corrections can conflate
actual photoconversion efficiency variations with (flux-conserving) pixel grid distortions that are produced by lateral electric fields within the CCD.  

\item{} Many (probably all?) of the unwanted effects we see in deep depletion CCDs are exaggerated versions of phenomena that afflict traditional CCDs.
Device thickness and wafer resistivity are continuous variables, and the phenomena described below surely exist at some level in all devices.  

\item{} The ``precision frontier'' in astronomical measurements will yield substantial scientific gains. We are nowhere near the Poisson limit for precise flux measurements when N$_e~>$1000, for example. 

\item{} The optical/IR astronomical instrumentation community seldom invests the time and effort needed to truly understand the nuances of their instruments.  Tendency for teams to produce a system, put it on the sky, and move on to writing the proposal for the next instrument. Science suffers as a result. 

\item{} It is incumbent upon those of us who design, build and deploy next-generation systems (including instruments, algorithms and methods) to identify and overcome 
sources of systematic errors that are an impediment to attaining high precision and accuracy in astronomical measurements. 

\end{enumerate}

\section{Lateral electric fields as a unifying framework, and establishing a common vocabulary}\label{sec:Efields}

The telescope produces a distribution of photon intensity $I_\gamma(u,v,z,\lambda)$ within the detector volume, where $u,v$, and $z$ are the spatial coordinates 
in the silicon, with $z$ parallel to the main optical axis. 
The eventual manifestation of this incident light as photoelectrons distributed across the discrete pixels of the CCD, $N_e(i,j)$,  is governed by
a combination of photoconversion efficiency $QE(u,v,z,\lambda)$ within the silicon as well as a charge transport matrix $J$ that determines where
the photoelectrons (or holes) end up after they are produced. After a time t the combination of photo conversion and charge transport eventually generates a monochromatic image 
given by
\begin{equation}
\label{eq:1}
N_e(i,j,\lambda)=\int QE(u,v,z,\lambda)~I_\gamma(u,v,z,\lambda)~J(u,v,i,j,\lambda,time,temperature,N_e(i',j',t)...)~\mathrm{d}u \mathrm{d}v \mathrm{d}z \mathrm{d}t,
\end{equation}

\noindent 
where the charge transport properties are encoded in a matrix $J$ (which describes the Jacobian mapping between the focal surface and the 
pixel array) that can in principle have a complicated dependence on 
time, wavelength, temperature and (for strain-induced internal electric fields) the thermal history of the system. Note the inclusion of the evolving 
charge distribution $N_e(i',j',t)$ as one of the parameters that governs the Jacobian. This nonlinearity can produce intensity-dependent
(and anisotropic) distortions to the PSF, and distortions in the photon transfer curve obtained using flats. 
Any departure from a simple rectilinear mapping $J$ from the photon distribution to the pixel array (due to lateral electric fields and other effects) 
will produce variations in the flux detected in each pixel. These variations can be mistaken for QE variations, and dividing by the normalized flat field will
introduce errors in the extracted shape, position and flux of objects. 

Some of this is very familiar to careful astronomers. Illumination corrections using rastered images of point sources has long been used to correct
for (typically low order, axisymmetric) optical distortions and for other flat-fielding errors. But the phenomena described here can occur on a much 
smaller spatial scale, and with abrupt transitions that are not amenable to low order WCS-like corrections. 

\begin{figure}[tbp] 
\centering
\includegraphics[width=6in]{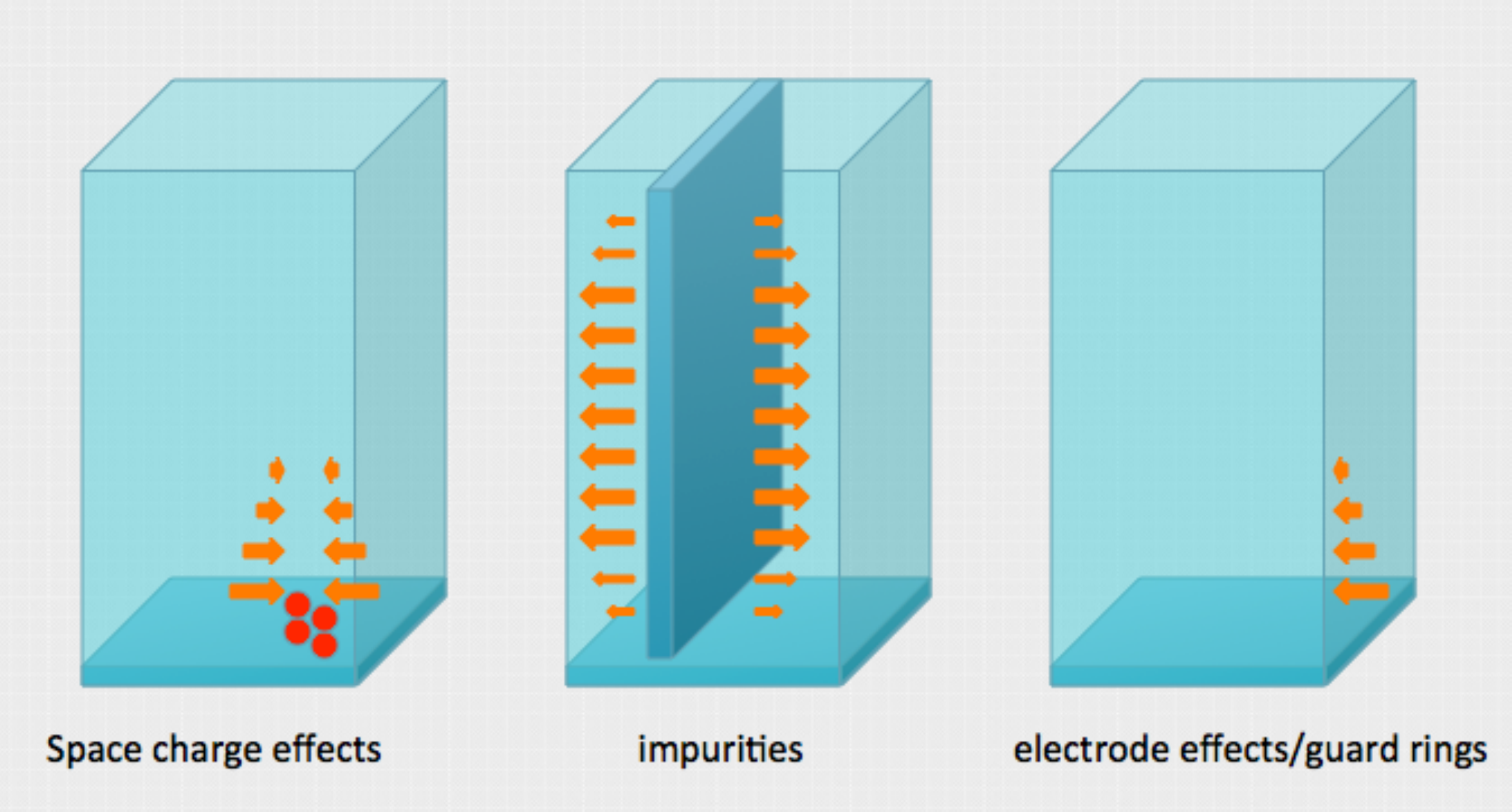}
\caption{Lateral Electric Fields in thick CCDs. This figure illustrates some of the phenomena that produce lateral electric fields within
CCDs, including (left) field and potential perturbations arising from gradients in accumulated photo-generated charge, (middle) imprinted 
fields due to impurity gradients in the silicon, and (right) lateral electric fields due to guard rings around the device periphery. Each of these
can produce charge transport effects that influence the 3-space to 2-space mapping between the photons in the detector volume and the 
manifestation of the resulting photocharge across the pixel array. It is essential that we be able to distinguish these charge transport 
perturbations from variation in quantum efficiency. Not doing so introduces errors during flat fielding. Moreover, these effects can violate common assumptions 
about a regular rectilinear pixel grid and intensity-independent point spread functions.}
\label{fig:Efields}
\end{figure}

Figure \ref{fig:Efields} illustrates some of the processes that can give rise to a non-zero charge transport Jacobian. Lateral electric fields
can perturb the trajectory of a charge after it's produced by photon absorption. Thicker CCDs are more susceptible to imperfections from 
these lateral fields simply because the charge spends more time making its way to the constraining pixel array, and it therefore suffers
a larger lateral displacement. The magnitude of wavelength dependence is governed by the z-dependence of these lateral fields. Table 
\ref{tab:Efields} lists some conjectures about the likely chromatic dependence of some of these processes. The following sections 
attempt to describe and define some of these effects in a bit more detail. 

\begin{table}[tbp]
\caption{Detector Features that Contribute to the Jacobian charge transport matrix $J$ within a CCD.}
\label{tab:Efields}
\smallskip
\centering
\begin{tabular}{|l|l|l|}
\hline
{\bf Phenomenon} & {\bf Length Scale} & {\bf Wavelength Dependence?}\\
\hline
Edge Distortion & 10-12 pixels around edges & Minimal?\\
Charge Correlations & $\sim$ 6 pixels & Minimal? \\
Impurity Gradients & Variable, radial symmetry & Yes \\
Lattice Strains & Varies & ?? \\
Pixel Fabrication Errors & Varies, some are periodic & Minimal\\
\hline
\end{tabular}
\end{table}

\subsection{Edge distortion effects}

Figure \ref{fig:edge} shows the perturbation in collected charge near the edge of an LSST prototype sensor, when subjected to uniform illumination. 
The contribution by O'Connor shows that scanning a spot of illumination up to the edge of the array yields a constant flux within an appropriate aperture, 
but the flat-field measurement of surface brightness 
clearly changes at the edges of the array.  We interpret this as a distortion in $J$, driven by lateral fields from the guard ring, that distort the effective size
of the pixels at the edge of the sensor. This is seen in the Dark Energy Survey (DES) sensors as well, but with the opposite sign. Their edges show enhanced
rather than diminished flux. The long coherence length of the electric fields makes this effect trivial to detect, and I propose we define this
phenomenon as {\em edge distortion}.  Henceforth suggested definitions will be in {\em italics}. 

\begin{figure}
\centering 
$
\begin{array}{cc}
\includegraphics[width=3.in]{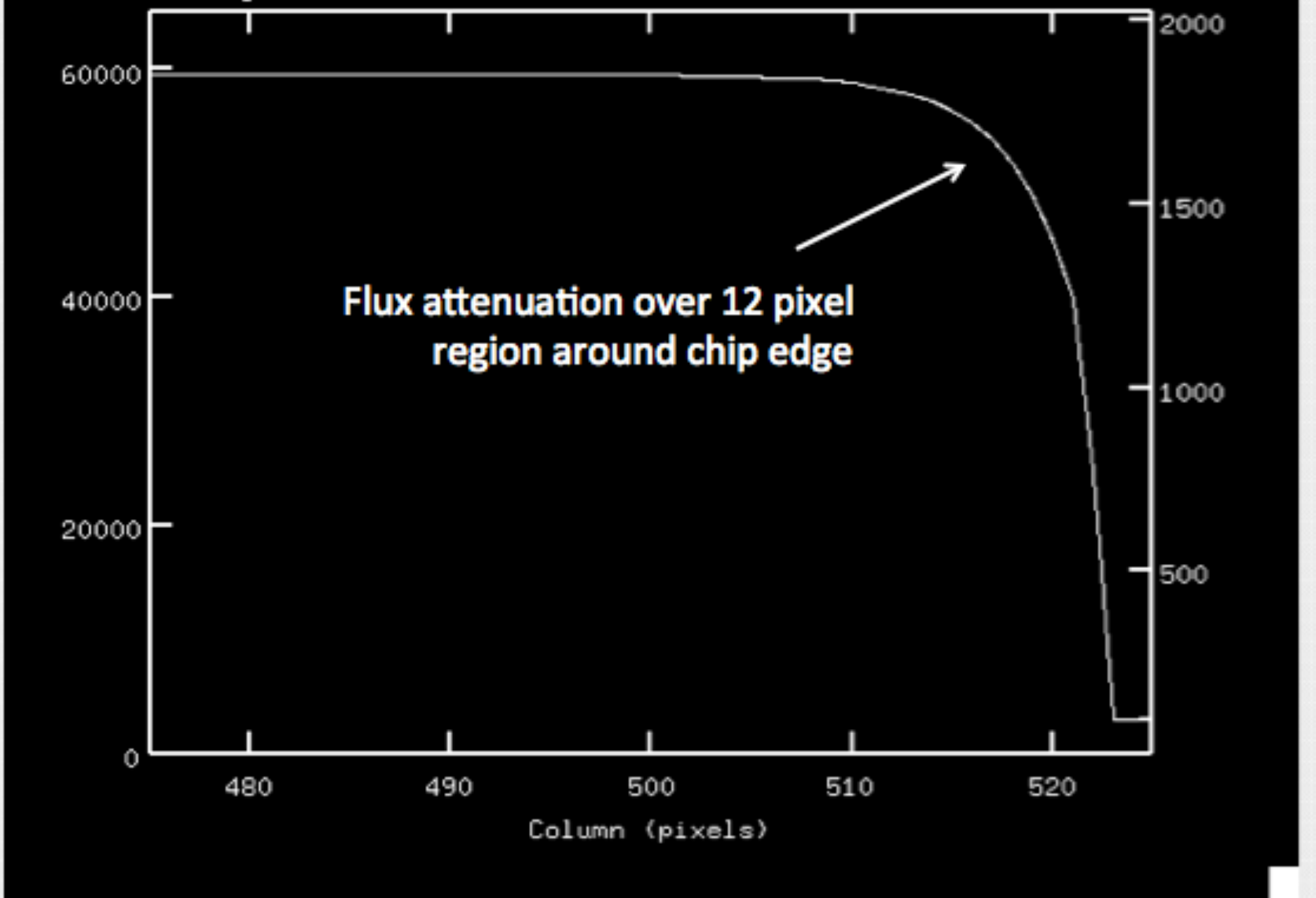} & 
\label{fig:edge}
\includegraphics[width=3.in]{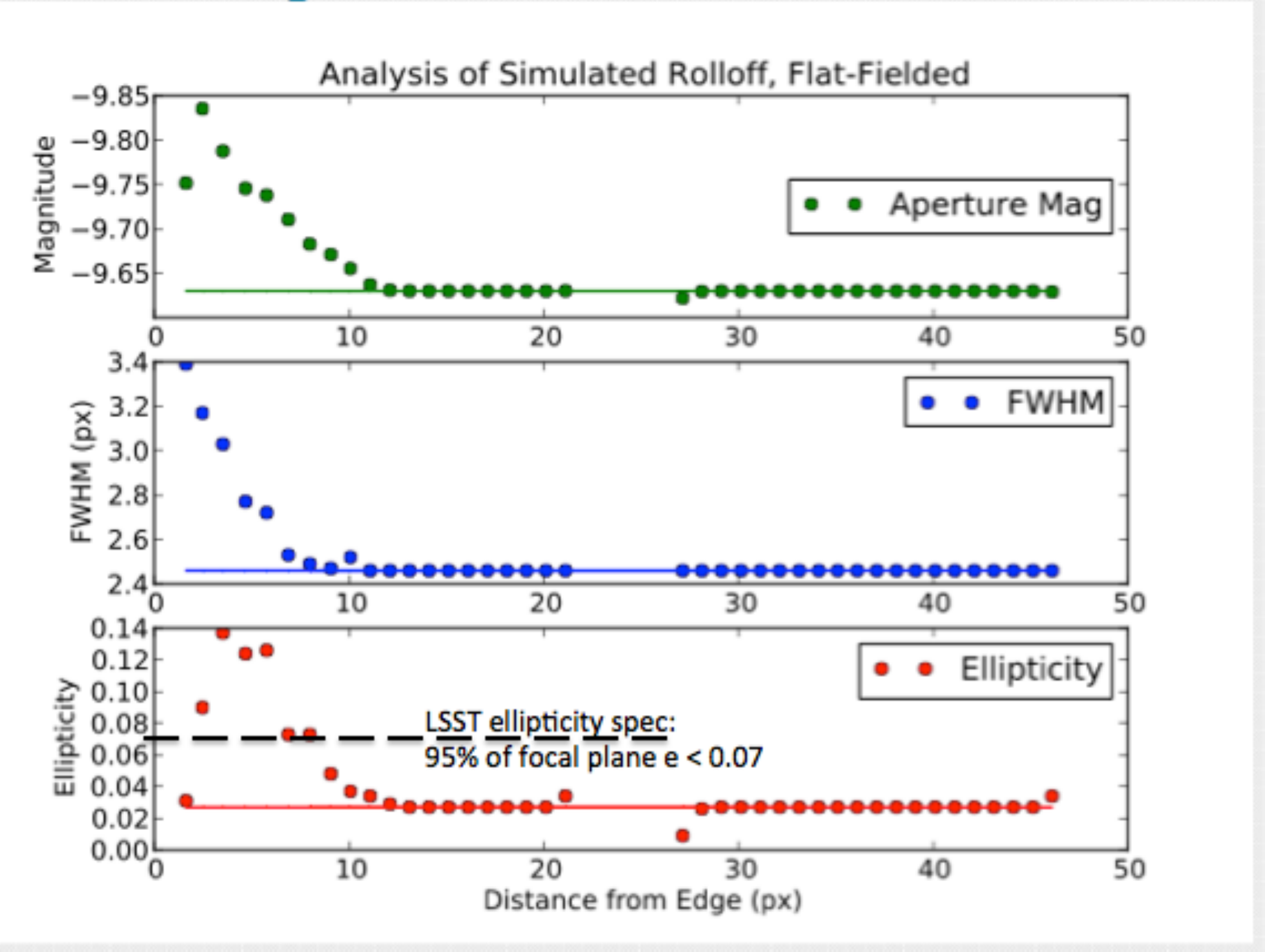}
\end{array}
$
\caption{Edge rolloff in an LSST prototype sensor. The CCD is 100 $\mu$m thick, and has a guard ring running around the perimeter of the sensitive area. 
Stray fields from the guard ring perturb the charge transport near the edges of the sensor. The left panel shows the flux rolloff under uniform illumination, 
starting about 12 pixels from the edge. The right panel shows the expected perturbations to the magnitude, position and shape of a star if the images
were mistakenly flat-fielded using the uncorrected attenuated flat. (Data from P. Doherty, Harvard LSST group, modeling by A. Vaz, also Harvard LSST group). }
\label{fig:edge}
\end{figure}

\subsection{Impurity gradients}

The fabrication method for the high-resistivity silicon boules from which thick CCDs are made introduces annular regions of varying impurities. Once the wafers are sliced
from the boule, and a CCD is made on the wafer, these impurity variations appear as concentric rings in flatfields. There are other aspects of the fabrication process
that can also produce similar-looking swirls, such as polishing, lapping, and backside processing. But the ``tree rings'' contain varying amounts of impurities, 
and the resulting spatial gradients in charge density produce lateral electric fields that are evident in astrometric and photometric residuals in both the PanSTARRS
and the Dark Energy Survey data. Since these impurity gradients run the entire vertical span of the sensor, the perturbations to astrometry and photometry 
do have a non-trivial wavelength dependence. Blue bands have a larger effect since a larger fraction of the photons convert near the back surface, and the 
resulting charges travel (on average) a larger vertical distance than do charges produced in redder bands.  Examples of this {\em impurity gradient}  phenomenon
are presented in these workshop proceedings by PanSTARRS (presented by Magnier) and DES (presented by Bernstein and Plazas). 

\subsection{Charge Correlation Effects}

Many of us believe that the observed bending of photon transfer curves and the intensity-dependence of the PSF are both driven by the repulsion of 
newly generated photocharge by previously accumulated charge distribution in the pixels (for an alternative interpretation see the contribution by S. Holland). 
These {\em charge correlation} effects can change the shape as well as the FWHM of the PSF, by an amount that depends on the detailed charge distribution in the 
CCD array. This is a potential source of numerous subtle errors in the measurement of galaxy shapes.  For example if bright stars at high SNR are used to estimate
the PSF shape in order to compensate the measured shapes of faint galaxies, we'd be making an error. In addition, the observed $\sigma$ is less
than the Poisson value of $\sqrt{N}$ due to correlations between the flux values in nearby pixels. We need to properly take this into account. This nonlinear effect
can't be corrected by convolution with some universal kernel.  Since these interactions take place close to the front (pixel) side of the device, we expect 
for there to be only a small wavelength dependence for the charge correlation coefficients. The contributions by Astier and Holland discuss these charge 
correlation processes. 

\subsection{Lattice stresses}

The DES sensors show evidence of lateral fields in the vicinity of the tape pieces that are used to hold down the CCDs while they are being bonded to the substrate. 
This indicates that we should be careful about any stresses that we introduce into the sensor, be it from packaging, dislocations introduced by sawing and 
back-side processing, or any other mechanical deformation or warping. The packaging stresses that are applied to detectors to ensure that they are flat 
enough to use in fast optical beams could well be producing lateral field structures over a variety of length scales. Depending on the vertical distribution of 
the stress gradients in the device, these {\em lattice stresses} could well introduce wavelength-dependent charge transport anomalies. 

\section{Pixel Fabrication Errors}

In addition to the variation in apparent pixel size produced by the lateral field effects summarized above, we do need to remember that not all pixels
are created equal. Fabrication errors can and do produce real physical differences in the sizes of pixels in CCDs, both thick and thin. Some of these 
arise from mask errors, and some of these effects are reproducible from wafer to wafer in a production run for CCDs that occupy the same place on the wafer. 
Other pixel area variations are more stochastic, for example those arising from meandering implanted channel stops. 

\subsection{Periodic mask errors}

\begin{figure}
\centering 
\includegraphics[width=5in]{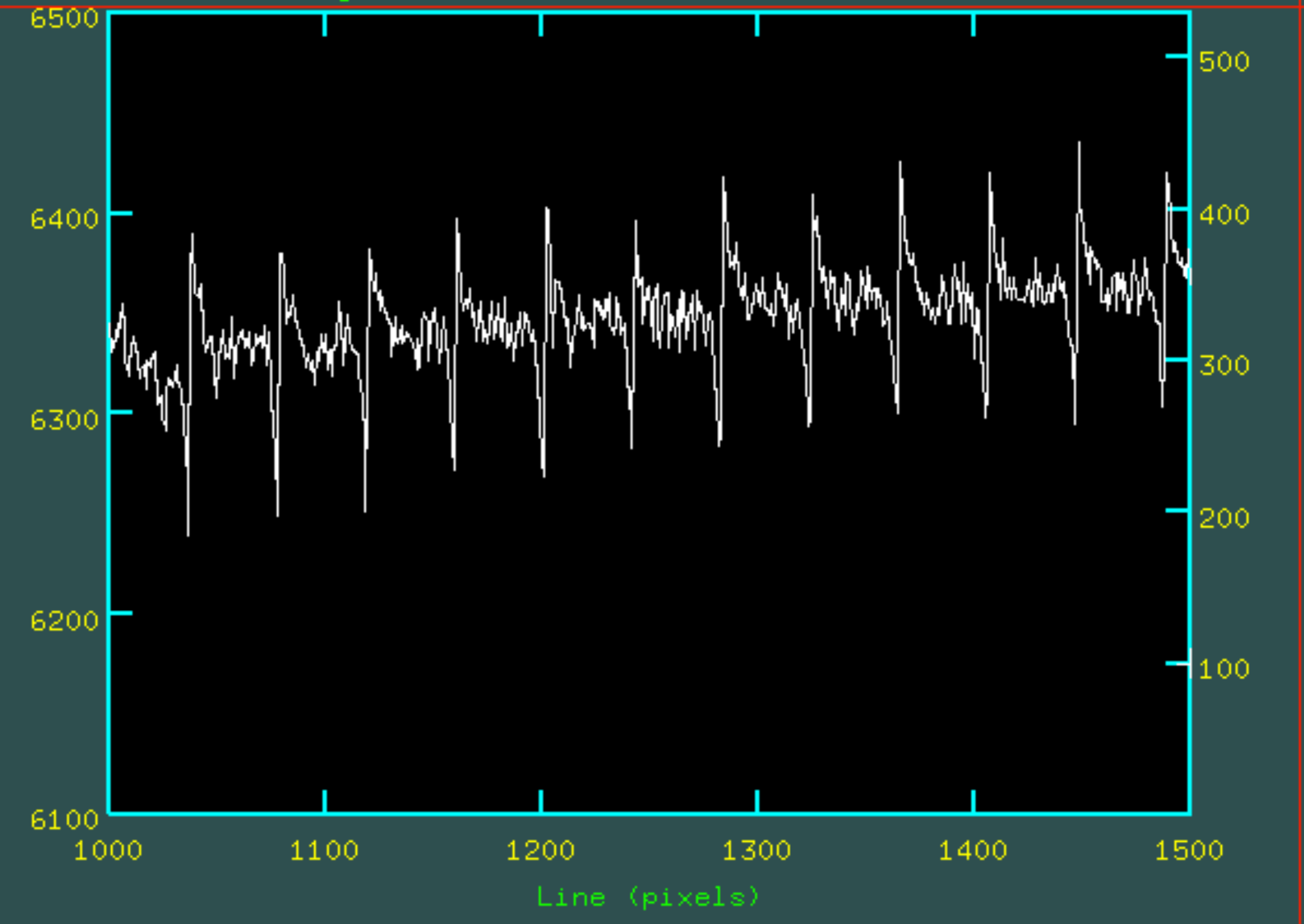} 
\caption{Periodic Mask Errors. This plot shows the result of averaging together 300 columns from a flat-field image. The resulting 
mean flux is plotted as a function of line number. The 41-row periodic structure is due to step-repeat errors in mask fabrication. 
Data taken by P. Doherty, Harvard, on an LSST prototype sensor. This source of physical pixel size variation is well known 
in the CCD instrumentation community.}
\label{fig:mask}
\end{figure}

Periodic mask errors are a well-known feature of CCDs, and we've typically either ignored them, or made things worse. 
Figure \ref{fig:mask} shows that some of the LSST
prototype sensors exhibit a periodic glitch in flat field response every 41 rows. This is another instance where traditional flat fielding produces an 
appropriate correction for a measurement of surface brightness (flattened sky looks flat) but we are introducing an error for the measurement of 
fluxes of celestial sources, since the total charge collected from a source does not care about the sizes of the pixels it illuminates. 
If such a detector is used in a spectrograph with the dispersion direction running down its columns, this {\em periodic pixel fabrication error}
can introduce periodic transients in the wavelength solution, {\it i.e.} the correspondence between pixels and wavelength, for algorithms
that assume a constant spacing between adjacent pixels.   The contribution to this workshop by Smith shows interesting FFT analyses of 
flat-field images that can be used to track down periodic features in CCDs. 

\subsection{Aperiodic pixel fabrication errors}

There are also sources of {aperiodic pixel fabrication errors}, and I think we know very little about these since with flatfields these pixel size variations 
are degenerate with sensitivity variations. A careful combination of laboratory and on-sky measurements of the fluxes and positions of sources, 
natural and artificial, will be very helpful in mapping out the spatial structure and correlations of these fabrication errors. Although it might be
very difficult to distinguish physical size variations from the apparent size variations driven by lateral electric fields, I'm not sure we that 
from a pragmatic perspective we need to discriminate between them.
 
\section{Conclusions}

After reaching this point in the document, the reader might consider it rather miraculous that CCDs work at all. But it's important to bear in mind that
we're talking about subtle effects, typically below 1\% for photometry, for example. 
The current generation of deep depletion CCDs are exquisitely high quality sensors. But as we strive to make the very most of the 
astronomical observations we obtain, we can improve our measurement quality by identifying and overcoming even subtle 
detector imperfections. It seems we are turning a corner in the exploitation of CCD data, and the initial attempts to correct for these
detector systematics are indeed producing unprecedented precision, even for wide-field surveys. 

I have come to the point of view that much of the structure seen in CCDs on small spatial scales, both thick and thin, that we have historically attributed to 
QE variations actually comes from pixel size variability, both physical and virtual (arising from lateral field effects). There are certainly actual 
variations in photon sensitivity, from fringing, variations in backside processing, and dust spots, but I am led to wonder how much damage
we have needlessly imposed on both astrometry and photometry, over these many decades!  

We seem to be just at the stage of arriving at a consensus (for the most part) on the list of features we have collectively identified in fully depleted, high-resistivity
CCDs. The next phases will presumably include 

\begin{itemize}
\item{} making device models of appropriate sophistication, in order to more fully understand these effects, 
\item{} devising laboratory measurements to map out the wavelength and position dependencies of the various phenomena,
\item{} continuing to use on-sky data to assess the perturbations caused to the determination of fluxes, shapes and positions of sources,
\item{} making a rank-ordered quantitative list of the effects of concern, for any given science objective, 
\item{} guided by this impact-ranked list, devise algorithms to avoid, dodge, overcome, suppress or correct for the effects that dominate our systematic error budgets
\item{} fold our improved understanding into making better detectors: consider architecture and fabrication changes that might yield CCDs with superior performance
\item{} fold our improved understanding into making better observations: optimize dither patterns and exposure times so as to overcome these effects, and
\item{} fold our improved understanding into improved instrument designs: optimize the sampling of the PSF and the optical design to take physical length scales into account. 
\end{itemize}

I have attempted to provide a context and unifying framework for the many interesting contributions to the 2013 BNL workshop
on Precision Astronomy with Fully Depleted CCDs, and on occasion to point 
out a few specific contributions that highlighted or illustrated some of the points raised in this narrative. But those of us who attended the meeting 
found each and every presentation to contain valuable information and insight, and so individuals who share my keen interest in 
pushing the boundary of astronomical precision are encouraged to study each of the valuable and interesting contributions in this volume.

%
%

\newpage 

\acknowledgments

My interest in the inner workings of deep depletion CCDs was stimulated by my work with the LSST sensor team. I am particularly grateful to P. Antilogous, P. Astier, 
P. Doherty, D. Finkbeiner, K. Gilmore, A Guynnot, D. Huang, I. Kotov, R. Lupton, G. Magnier, A. Nomerotski. P. O'Connor, A. Rasmussen, N. Regnault, S. Ritz, J. Tonry and A. Vaz for helpful and (for me) educational conversations.

I am also grateful to the technical teams at e2v and the University of Arizona Imaging Technology Laboratory, and to the 
workshop participants for illuminating and stimulating presentations and discussions. A special thanks to Andrei Nomerotski and his BNL 
colleagues for arranging a workshop that was most timely, interesting and informative. My own understanding of these sensor issues 
evolved considerably during the course of the meeting, and some of that increased understanding is reflected in this narrative. 

My work on these topics is supported in part by the US Department of Energy under grant DE-SC0007881, by STSCI under grant HST-GO-12246.01, and by the DOE's LSST Camera Project.


\begin{thebibliography}{9}

\bibitem{WASP}
J. Tregloan-Reed \& J. Southworth, {\em An extremely high photometric precision in ground-based observations of two transits in the WASP-50 planetary system},
MNRAS {\bf 431}, 996 (2013).

\bibitem{Astier} 
P. Astier et al., \emph{Photometry of supernovae in an image series: methods and application to the supernova legacy survey (SNLS)}
A\&A {\bf 557A}, 55A (2013).

\bibitem{WFPC}
J. Holtzman et al., \emph{The performance and calibration of WFPC2 on the Hubble space telescope}, PASP {\bf 107}, 156,(1995).

\bibitem{Kotov}
I. Kotov et al., \emph{Study of pixel area variations in fully depleted thick CCD},
Proc. SPIE {\bf 7742}, 774206-1 (2010). 

\bibitem{Smith}
R. Smith \& G. Rahmer
\emph{Pixel area variation in CCDs and implications for precision photometry}, 
Proc. SPIE {\bf 7021}, 70212A-1, (2008). 

\bibitem{Downing}
M. Downing et al., 
\emph{Bulk silicon CCDs, point spread function and photon transfer curves: CCD testing activities at ESO},
{\emph{Proc.\ SPIE.} {\bf 76} (2005) 054503}

\end{thebibliography}
\end{document}